\documentclass[MSNbibl,number,dvips]{arxstspdf}
\usepackage{flushend}
\usepackage{stfloats}
\usepackage{graphicx}
\volume{27}
\issue{4}
\pubyear{2012}
\firstpage{594}
\lastpage{606}
\doi{10.1214/12-STS388} 

\setattribute{abstract}{width}{348pt}
\setattribute{keyword}{width}{348pt}

\begin{document}
\begin{frontmatter}

\title{A Conversation with David Findley}

\runtitle{A Conversation with David Findley}

\begin{aug}
\author[a]{\fnms{Tucker S.} \snm{McElroy}\ead[label=e1]{tucker.s.mcelroy@census.gov}}
\and
\author[b]{\fnms{Scott H.} \snm{Holan}\corref{}\ead[label=e2]{holans@missouri.edu}}

\runauthor{T. S. M\lowercase{c}Elroy and S. H. Holan}

\affiliation{U.S. Census Bureau and University of Missouri}

\address[a]{Tucker S. McElroy is Principal Researcher, Center for Statistical Research and
Methodology, U.S. Census Bureau, 4600 Silver Hill Road, Washington, DC
20233-9100, USA
\printead{e1}.}
\address[b]{Scott H. Holan is Associate Professor, Department of Statistics, University of Missouri,
Columbia, Missouri 65211-6100, USA \printead{e2}.}

\end{aug}

\begin{abstract}
David Findley was born in Washington, DC on December 27, 1940. After
attending high school in Lyndon, Kentucky, he earned a B.S. (1962) and
M.A. (1963) in mathematics from the University of Cincinnati. He then
lived in Germany, studying functional analysis under Gottfried K\"othe,
obtaining a Ph.D. from the University of Frankfurt in 1967.  Returning
to the United States, he served as a mathematics professor at the
University of Cincinnati until 1975. Having transitioned from pure
mathematics to statistical time series analysis, Findley took a new
academic position at the University of Tulsa, during which time he
interacted frequently with the nearby research laboratories of major
oil companies and consulted regularly for Cities Service Oil Company
(now Citgo).  In 1980 he was invited to lead the seasonal adjustment
research effort at the U.S. Census Bureau, and eventually rose to be a
Senior Mathematical Statistician before his retirement in 2009. In 1966
he married Mary Virginia Baker, and they currently live in Washington,
DC.

David Findley has published more than 40 journal articles and book
chapters, as well as dozens of technical reports and conference
proceedings, many of which are heavily cited and influential.  He has
also published two edited volumes (1978 and 1981) that have had a
substantial impact on the field of time series analysis.  Numerous
honors and awards have accrued to him, including ASA Fellow (1987), the
Julius Shiskin award (1996) and the U.S. Department of Commerce Gold
Medal (1997).
\end{abstract}

\vspace*{-3pt}

\end{frontmatter}

\begin{figure}

\includegraphics{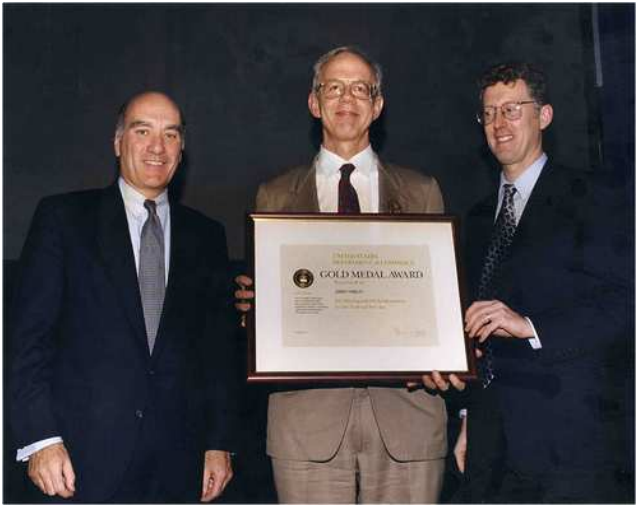}

 \caption{David Findley receiving the Commerce
Gold Medal 1997.  David Findley was honored with this award for
scientific leadership and contributions to the field of time series
analysis, especially seasonal adjustment research.}\label{f1}
\end{figure}

The initial conversation between David Findley, Scott Holan and
Tucker McElroy took place on July 13, 2010 at the U.S. Census
Bureau.  Holan and McElroy later obtained clarifications of certain
points from David Findley during the following year.

\section{Education}

 \textbf{McElroy}: Hello, David. We are thankful to have the opportunity to
discuss your life and career.  Could you describe for us the early
influences that led you to pursue a career in mathematics?

\textbf{Findley}: In high school I~found algebra and geometry
enjoyable and interesting.  I~found out decades later that many of
my relatives were mathematics teachers, including one who taught in
Suitland High School, not far from the U.S. Census Bureau (USCB).
However, my plan was to study physics, and I~decided to become a
mathematics major only when I~found out that the mathematics
department at the University of Cincinnati would let me take
advanced calculus and linear algebra before completing the basic
calculus sequence, whereas the undergraduate program chair of the
Physics Department would not accept the summer version of the
general physics course as an adequate prerequisite for the
upper-level physics courses that I~wanted to take my sophomore year.
By majoring in mathematics, I~was able to take the physics courses I~wanted, including a graduate level electrodynamics course my senior
year, along with enough graduate mathematics courses that I~was
prepared to complete the master's program in mathematics a year
later.

\textbf{McElroy}:  I~see; but this was in Cincinnati, whereas you were
born in Washington, DC.

\textbf{Findley}: I~spent the first 12 years of my life in Washington
DC, but moved with my family to Dayton, Ohio, which is close to
Cincinnati.

\textbf{Holan}: So your interests in mathematics primarily began in
high school?

\textbf{Findley}:  No, I~was interested in physics.  I~mean, I~enjoyed mathematics, but I~was set on physics from an early age.
Perhaps because I~had been very impressed by the age of four or five by
the atomic bomb, having seen pictures of it on the front page of the
newspaper.

\textbf{Holan}: That leads to our next question: why you chose the
University of Cincinnati.  Did you consider any other schools, such
as The Ohio State?

\textbf{Findley}: I~only looked at the University of Cincinnati (UC),
because a high school counselor had recommended it.  He knew that
its College of Engineering had a strong reputation, and assumed that
its Physics Department would be equally strong, which it wasn't at
that time.

Beyond my getting a solid mathematics background, a few other
circumstances there also had a large impact.  The Math Department's
flexibility continued into graduate school.  Because of my interest
in physics, I~was interested in Hilbert space theory.  There was no
expert in functional analysis on the faculty, but I~and two more
senior graduate students were allowed to give ourselves a reading
course in the area with nominal faculty supervision, working our way
through the masterful monograph by Riesz and Nagy \cite{RiNa55}.

Then I~was permitted to write a master's thesis in the area on the
topic I~chose,  expositing a Russian mathematician's paper on
operator representations, generalizing the representation formulas
for symmetric and unitary operators, which generalize the
representation formulas for symmetric matrices and unitary matrices.
As a consequence, the first time I~encountered the backshift
operator in time series analysis and the Cram\'er integral
representation of stationary processes, I~recognized them as special
cases of unitary operator representations.  This led me to conclude
that I~had an advantageous background for time series analysis,
which led me (years later) to choose this as my area of statistics.

My contacts with Germany also came from UC. Freshman physics majors
were required to take German, and---perhaps because of Cincinnati's
German heritage---there were outstanding teachers in the German
Department.  After three semesters of German courses, I~felt I~had a
solid foundation for studying in Germany, when later the idea came
to me of combining my desire to live in Europe with my desire to get
a Ph.D. in mathematics.

Fortunately, for me, the Mathematics Department in Cincinnati had a
professor from Germany named Arno Jaeger, who knew the functional
analyst in Germany I~had wanted to work with, Gottfried K\"othe in
Heidelberg.  Jaeger also knew a German mathematician at the
University of Maryland in College Park, who could offer me a
research assistantship after I~learned that K\"othe would be in
College Park for the 1963--1964 academic year.

\textbf{McElroy}: After the University of Cincinnati, you held a
research assistant position at the University of Maryland.  What
sort of research were you doing at the time, and did this have any
influence on the problems you would later consider in your career?

\textbf{Findley}: The research assistant position had variable duties
depending on who you were working for.  I~marked test papers for a
first year graduate course and proof-read a Ph.D. dissertation.  At
Professor K\"othe's request, I~presented an exposition of a recently
published paper by two Japanese mathematicians in the Functional
Analysis seminar series that ran for decades in the home of
Professor John Brace in College Park.  Their paper provided a
complicated counter-example to a long-standing open question.  Later
in Germany, I~learned that seminar presentations like this were the
main filter used by professors to select Ph.D. dissertation
students.

\textbf{Holan}: When you were at the University of Maryland (UMD), were there
any young professors there with whom you came into contact?

\textbf{Findley}: There were two graduate students who showed the
rest of us what it was like to be really gifted! Simon Levin and James
Yorke; James has become very well known for his work in chaotic systems
theory.  He was a graduate student there, and he was then hired after
graduation by UMD.  I~know the time series analysts at Maryland, but
they came later, and I~got to know them by giving seminars at the
department.  That is where I~met my future co-authors Ching-Zong Wei
and Benedikt P\"otscher, who later left UMD---but who were there at the
time that I~started giving seminars.

\textbf{Holan}: At the time you were a Ph.D. student, was it common to go
study abroad?

\textbf{Findley}:  My parents had traveled in Europe, and seemed
to enjoy it a lot, and I~felt that my German language training was
solid enough that I~probably would survive.  In Heidelberg there
were a number of foreign students, and also other mathematics
students from the United States.  But, I~was the first US citizen
to get a doctorate from the Faculty of Natural Sciences of the
University of Frankfurt.

\textbf{Holan}: Were your classes in German or in English?

\textbf{Findley}: German; keep in mind that mathematics is probably
the easiest subject to study in a foreign language.  The vocabulary
is pretty limited and pretty predictable in its forms!

\textbf{McElroy}: It seems that during the sequence of your studies, there
was an interval of time between Heidelberg and Frankfurt, at the
University of Cincinnati.

\textbf{Findley}: Yes.  I~came back to the US for various personal
reasons, including a death in the family, and also to earn enough
money to marry my fianc\'ee.  Also, I~wanted some time to decide
whether I~wanted to continue in mathematics, or return to physics.

\textbf{McElroy}: Your Ph.D. work in Germany lasted three years.  What
was your dissertation about?

\textbf{Findley}: It concerned the study of a quite abstract
generalization---given in papers in 1963 and 1964 by W. A. J.
Luxembourg and A. C. Zaanen---of the general class of vector space
of real-valued sequence spaces defined by K\"othe and his Habilitation supervisor Otto Toeplitz in a 1934 paper.  I~was able to prove analogs, in
this new abstract setting, of several results established by K\"othe
in a series of papers from 1935 to 1951, results that made these
sequence spaces quite influential in the development of the theory
of locally convex topological vector spaces.  The only part of this
work I~have been able to use in my time series research is a
characterization I~learned then of compact sets in the vector
space of absolutely summable sequences.  This played a minor
clarifying role in a later paper with Benedikt P\"otscher and
Ching-Zong Wei on almost stationary processes \cite{FinPotWei01,FinPotWei04}.

\textbf{Holan}: So, you decided to go back to the University of
Cincinnati after graduation?  Was this your first job out of school?

\textbf{Findley}: Yes, I~felt it was an honor to be invited back.  I~knew and liked the department, and also my wife, who is a violinist,
wanted to continue studying with the violin teacher she previously
had at the College-Conservatory of Music at the University of
Cincinnati.  So it suited both of us.

\textbf{Holan}: So, at the time they just invited you back to apply?
Today it seems to be much more of an adventure finding a job out of
school.

\textbf{Findley}: It was a slightly different time.  Perhaps a
year or two after I~was back at the University of Cincinnati, the
job market for new Ph.D.s in mathematics became quite tight.
However, there were many jobs available at the time I~applied.  In
fact, I~was also encouraged to apply for a position at the
University of Maryland; so, I~wasn't worried about opportunities in
mathematics.

\section{Transition to Statistics}

\textbf{Holan}: Your return to Cincinnati seems to mark the time you
transitioned to time series analysis.  Given your physics interests and
the inherent  dynamics of many physical processes, it seems to be a
natural development.  Where did you learn time series analysis?

\textbf{Findley}: I~was self-taught.   I~found that when I~began
writing research papers in mathematics that the audience, perhaps
because of the area in which I~wrote my dissertation,  was very small
indeed and that the\vadjust{\goodbreak} amount of effort necessary to write papers that
would get into respected journals was so great that I~became somewhat
frustrated. Additionally, I~felt that most mathematicians looked to the
mathematics literature for stimulation for research, rather than
anything outside \mbox{mathematics}. Some were even contemptuous of the
mathematical work done in, say, econometrics and other disciplines. Now
in Frankfurt, through singing together in a chorus, I~had become
friends with the young Professor of Stochastics, Hermann Dinges (each
Full Professor had his own Institute then), with whom I~had taken
advanced seminars in Markov processes; he had suggested a research
problem to me involving Wiener--Hopf factorization. So I~was interested
in stochastics, and therefore also in statistics.  So after a year or
two I~started teaching statistics courses for engineering students, and
knew of the book by Box and Jenkins \cite{BoxJen76}.

In 1973, at the invitation of the sole mathematical statistician in
UC's Mathematics Department, Manny Parzen and Grace Wabha gave a one
day overview of time series analysis.  From this overview I~recognized
that time series play an important role in many disciplines, and that
the field uses mathematical tools that other statistical fields do not.
I~decided to become a time series analyst, and to start by reading
books with a probabilist colleague who found the one day overview very
stimulating.  Grace Wahba recommended Ted Anderson's 1971 book to us \cite{And71},
which we worked through very thoroughly.

Later, after I~accepted an offer for a time series position at the
Mathematics Department of the University of Tulsa (TU) in the Spring of
1975, I~was asked to prepare a course for the fall semester based on
David Brillinger's 1975 monograph \cite{Bri01}.  After reading his powerful
presentation of the frequency domain perspective, I~decided instead to
give a semes\-ter-long series of lectures for Mathematics Faculty members
on its contents, while presenting a more elementary two-semester course
for the graduate students, many of whom worked in exploration
seismology for local oil companies.  I~taught this two-semester course
every year at TU, updating it annually with new material, such as
state--space methods.

\textbf{McElroy}: What were your reasons for moving to TU?

\textbf{Findley}: Oh, they advertised the position in time series
analysis, and that was what I~wanted to do.  Also, they already had
three mathematicians on the faculty who had trained themselves in time
series analysis---perhaps\vadjust{\goodbreak} through consulting work for the oil company
research labs in Tulsa. So, I~knew I~would have a community of people
to work with. Additionally, they were willing to hire someone with no
publications in time series analysis, because they were aware that a
mathematician could become a competent time series analyst.

\textbf{Holan}: In addition to your colleagues, you also worked as a
consultant at Cities Service Oil. Can you tell us a little about your
experience there?

\textbf{Findley}:  I~should back up and mention that every mathematics
professor there did consulting work, because it was available.  The
university thought it was a perfectly reasonable thing, as long as they
did some other academic things. There were several major oil company
research labs in Tulsa and neighboring cities, and oil companies have
been quite at the forefront of using new technologies. They were among
the first to use radioactivity as a measuring tool, using gamma rays in
well logs to discover what was down there. So, there were interesting
scientific problems to work on and, in the particular case of Cities
Service, the university had been given a building which had belonged to
the Carter Oil company. Von Neumann had consulted in this building, and
it housed on one end most of the engineering college, and at the other
end the Cities Service research laboratories, which the university
leased to Cities Service. So I~walked out the door at one end of the
building and in the door of the other end, to consult, which was a very
nice arrangement!

Also, I~helped a geophysicist who was applying state space filtering
methodology---Kalman filtering---to some geophysics problems. I~worked
on some problems related to time series methods and seismology. There
were pure statistical issues---pure time series issues of a certain
kind, arising from physical reasons. If you calculate wave propagation
through a variety of media and assume everything is happening in one
dimension, autoregressive processes seem very natural, since partial
autocorrelations represent reflectivity coefficients, and anything that
is forecastable represents noise, usually arising from echoes of a wave
hitting the transition between geologic strata. Thus, anything that was
predictable was noise, and hence prediction error filtering was
important. I~developed some time-varying methods for predicting fifty
steps ahead---we were getting perhaps twenty-five hundred measurements
a second in these seismograms, so fifty was a useful forecast interval.

\textbf{Holan}: Did any of that consulting work generate mathematical
research?\vadjust{\goodbreak}

\textbf{Findley}: The paper on my algorithm for time-varying
forecasting for forecast error analysis was published in one of the
symposium proceedings volumes from the two symposia I~organized in
Tulsa \cite{autokey10,Fin81,autokey11}. It had a different approach to looking at Levinson's algorithm \cite{Lev47}.
(Mostly known to statisticians through the special case independently
discovered by Durbin~\cite{Dur60}.) I~was able to give a geometrical interpretation
of it, which seemed possibly new at the time.

 \textbf{McElroy}:  Can you talk a bit about those symposia?

\textbf{Findley}:  Yes, they were a great opportunity. My colleague J.
B. Bednar suggested, after I~had been at TU for a month or two, that I~organize a symposium on time series analysis. There were now four
people in the department interested in the subject, and I~saw this as
an opportunity to get in contact with the leading time series
researchers in a number of different fields. This was an idea that was
very exciting to me. I~asked J.~B.~Bednar to contact a very respected
electrical engineer at MIT, Alan Oppenheim, and ask if he would be
willing to speak at the conference. Once he received an affirmative
answer, I~contacted Manny Parzen, and after he said yes, I~contacted
Henry Gray. After that we were launched, because it began to look very
credible!

Akaike happened to be on sabbatical leave at Harvard; so I~contacted
him and he agreed to come. I~wound up inviting people from geophysics,
electrical engineering, exploration seismology---or seismology in
geophysics---mathematics, astronomy, and sta\-tistics. I~later discovered
that there had been a very successful symposium in 1962 that Murray
Rosenblatt organized at Brown University, whose proceedings had been
published as a book by Wiley, and that the Tulsa symposium was the
first thing like that, bringing people together from so many different
fields.  Therefore, there was really great interest in having such a
symposium. Additionally,  the first symposium was successful enough
that, when three years later I~organized another such symposium, things
went very well from the beginning!

\textbf{McElroy}: Were there any econometricians at the symposium?

\textbf{Findley}: Yes: Clive Granger and Rob Engle, who shared the Nobel
prize for economics in 2003. They gave very interesting talks, and
were delightful, very well-informed and interesting contacts for
later work.

\textbf{Holan}:  Did you keep in contact with many of the people at the
symposia?

\textbf{Findley}: Yes, indeed.  When I~later left TU and came to the
Census Bureau, I~lost contact with people from the geophysics and
seismology community, but I~certainly maintained contact with most of
the statisticians who came.  In statistics, there was Manny Parzen,
Hirotugu Akaike, Henry Gray, G. S. Watson, Richard H. Jones, Doug
Martin, David Donoho, William Dunsmuir, Will Gersch, Genshiro Kitagawa,
Wayne Woodward and Mel Hinich.  The electrical engineers Alan
Oppenheim, Thomas Parks, John Makhoul and Jerry Mendel also attended,
as well as the geophysicists Enders Robinson, Sven Treitel and Freeman
Gilbert.

Will Gersch later came to USCB as an ASA/NSF-Census research fellow,
with Genshiro Kitagawa as his research associate.  Akaike, Genshiro
and I~worked together when I~visited the Institute of Statistical
Mathematics in Tokyo.  My contact with Akaike was very long-lasting
and very fruitful for me. Just about everyone I~mentioned is someone
I~have benefited substantially from at later times, through
contacts of one kind or another.  Also Enders Robinson can be
classified as both a time series analyst and a geophysicist; he has
written a number of time series books \cite{Rob80}. His prediction error methods
were very important for exploration seismology.  Freeman Gilbert was
there. Freeman is a very well-known seismologist whom David
Brillinger has collaborated with.  Sven Treitel, a leading
exploration geophysicist, was there from Amoco.  So it was a stellar
experience for me to interact with these people!

\begin{figure*}

\includegraphics{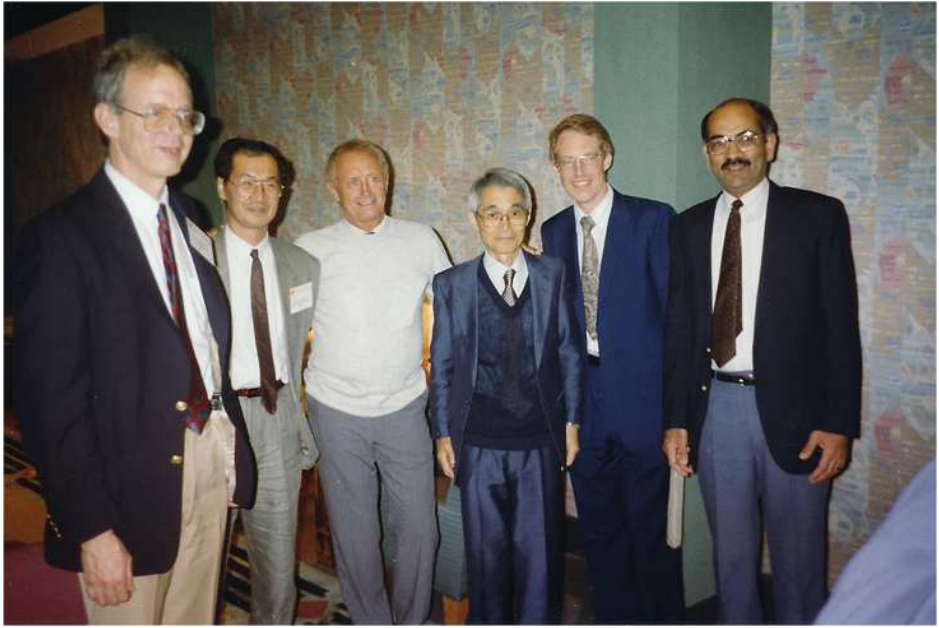}

\caption{David Findley, Kunio Tanabe, Will Gersh,
Hirotugu Akaike, Wallace Larrimore, and Raj Bhansali. (US/Japan
Conference on the Frontiers of Statistical Modeling: An Informational
Approach, held at the Department of Statistics, University of
Tennessee, Knoxville, to commemorate Akaike's 65th
birthday.)}\label{f2}\vspace*{-6pt}
\end{figure*}

\section{At the Census Bureau}

 \textbf{Holan}:  What led you to leave academia for a federal career?
Maybe you could tell us a little about what the Census Bureau was
like; for example, what the research environment was like and how
your transition from academia transpired?

\textbf{Findley}:  I~knew of the Bureau's reputation in the field of
seasonal adjustment and the influence of its X-11 software.  This gave
me the sense that any research done by me or my group would be noted
and considered by statistical offices and central banks around the
world---a larger potential audience than exists for most academic
papers. Also, I~was sure that there would be support for implementing
results of research in software designed for public use, making it much
more likely that the results would be used in practice.  This is in
contrast to the situation with software developed in academia, which
usually can only be\vadjust{\goodbreak} used by its author.  Thus, I~felt I~could have a
greater practical impact by working at USCB than by staying in
academia.

Finally, I~knew from having lived there that the Washington, DC area
was intellectually stimulating in the field of statistics, because
there were a large number of statisticians there.  I~was looking
forward to working with Bob Shumway at George Washington University
(GWU), but he was on sabbatical leave at the University of California,
Berkeley, and then accepted a position at Davis. So, Kirk Wolter of
USCB wrote to a number of prominent time series analysts, two of whom
recommended me to him. He then contacted me, and I~came to the Bureau
for an interview.

\textbf{McElroy}: It is interesting that your answer to that question
is almost identical to the answer that Agus\-tin Maravall gives for
why he chose a career at the Bank of Spain, namely to develop and
maintain a major software package.

\textbf{Holan}: So, in the early development of the software that came
out of Census, were you the original programmer, or did you oversee
the programming?\looseness=1

\textbf{Findley}: No, no;  I~was not one of the programmers. They had a
programmer already hired when I~came here.  He was someone with a
Master's degree in mathematics, but he had no background in
statistics, and therefore didn't really understand what he was
programming.  I~was lucky,\vadjust{\goodbreak} very early on, that Brian Monsell showed
up.  Brian has outstanding programming ability, and also a keen
interest in making software usable.  Brian was able to work easily
with lots of different operating systems and write good code, which
programmers at SAS and other people who have translated his code
have complimented.  So, one needs good luck, and Brian's coming
along was good luck, as was Bill Bell's being here, at the time!

Bill supervised much of the programming having to do with time
series modeling.  But, we didn't start out programming.  I~mean, we
started out just trying to respond to whatever the needs seemed to
be.  Our ultimate goal was to improve the practice of seasonal
adjustment at the Census Bureau and elsewhere.  USCB was still using
X-11 then, and hadn't even made the switch to X-11-ARIMA, so working
on that transition was important, and other things came along that
also seemed important.  Later we realized that there were quite significant
things that X-11-ARIMA couldn't do, namely estimate regression functions jointly
with the ARIMA models so that you could estimate
level shifts and other outliers, as well as other kinds of
user-defined regressors for special effects.

\textbf{Holan}: Given the production needs at the Census Bureau, how much time
did you spend doing research?  Was it your responsibility to ensure
that the research have an influence on production?\vadjust{\goodbreak}

\textbf{Findley}: Yes, it was my responsibility.  By way of background,
in 1978 and then in 1980, the Census Bureau had held two large seasonal
adjustment conferences with many distinguished speakers, including
Clive Granger.  As part of a way of building up momentum for a new time
series center at the Census Bureau, they had hired two
statisticians---one Master's degree statistician and one Ph.D. degree
statistician---in addition to the programmer I~mentioned earlier, and
already had them working. Kirk Wolter, who knew time series analysis,
as well as sample survey methodology, had been directing some work. So,
when I~arrived, I~had to learn about seasonal adjustment myself!  I~knew that it existed as a subject and that it had connections with
signal processing, but I~had a learning curve to go through.

But to go back to your question: Kirk Wolter had established monthly
meetings in which all the people in the Census Bureau interested in
seasonal adjustment attended, and some topic that seemed to be of
general interest was discussed.  I~found out, perhaps in the first
week, that Wayne Fuller, who did consulting for the Census Bureau, had
said that he would not work on seasonal adjustment until they started
doing concurrent seasonal adjustment, namely using all of the data
available every month, rather than forecasting seasonal factors from
December of the year before. Thus, I~had a mandate right away to make a
strong case for concurrent seasonal adjustment. These monthly meetings,
where the divisions discussed problems of interest to them, also
provided an opportunity where we could respond to their problems; we
could also present the research results we were obtaining on the
smaller revisions you got to seasonal adjustments---smaller revisions
after you recalculated the adjustment using later data---that came for
most of the time series when you used all of the data available, rather
than using forecasted seasonal factors from the December before.

This kind of idea, that
you shouldn't just use projected seasonal factors, had been a big
stimulus to Estela Dagum to introduce ARIMA models to forecast
ahead, so that you could at least use forecasted values as
replacements for the data that you weren't ready to use, or didn't
even have if it went past a certain point.  So, these monthly
meetings, which continued for several years, built up a rapport with
the other groups in the Census Bureau. I~think they may have gone on
for ten years.

\textbf{Holan}: In those times, what was the day-to-day process like in
terms of research\vadjust{\goodbreak} versus production? For example, suppose you are
sitting at your desk and you figure something out, on the topic of
concurrent adjustment, and you say, ``This really looks like what we
need to do.'' What was the process for getting things implemented in
production back in those days?

\textbf{Findley}: It was a matter of talking about it, and presenting
results in seminars. Also taking series from each one of the production
groups there, and giving them the results for each one of their groups.
Also, there were other things to talk about. The next big project that
came along was due to someone who, without consultation or approval,
seasonally adjusted the January value of a very important series,
petroleum imports, that hadn't been seasonally adjusted before because
it didn't exhibit stable seasonal characteristics. This happened when
Shirley Kallek, the Associate Director for Economic Statistics, was out
of town, and she decided that the published adjusted number had to be
replaced with the unadjusted number. This indicated a different
direction, which was embarrassing to the government. So she asked us to
develop diagnostics to correctly ascertain when series were not good
candidates for seasonal adjustment. That set off one of the veins of my
work that has continued until I~left the Census Bureau, and is still
ongoing. So, diagnostics and model selection are the two most
continuous veins of research that I~have pursued during my years at the
Census Bureau.

\textbf{Holan}: So---unusually for a federal employee---\break you've been
able to visit numerous different places, and work with many different
people, and spend a considerable amount of time away from the Census
Bureau, sort of like an academic on sabbatical.

\textbf{Findley}: Yes;  the thing I~was most nervous about in leaving
academia was giving up sabbatical leaves, because I~had already had one
and found it very fruitful to go away for a number of months to another
place, and interact with other people and share an office with someone
doing something completely different, but something that turned out to
be interesting to me. It was broadening in some way. It is the
international nature of seasonal adjustment work that to some extent
led to my receiving invitations to go to other places. It was also Bill
Bell's presence here on the time series staff, being perfectly capable
of taking over for me when I~was away, which was probably more
important in making it possible.

These visiting positions have always been productive for me, usually
enabling me to complete research related to model selection that I~couldn't find time for at USCB. I~have been especially fortunate to
have more than one such visit to the Institute of Statistical
Mathematics (ISM) in Japan and the Institute of Statistical Science of
Academia Sinica in Taiwan. Both have strong research groups in many
areas of statistics, superb libraries and other research resources, and
contacts with other organizations, including statistical offices
interested in seasonal adjustment.

Genshiro Kitagawa has always been my host at ISM, where Akaike had
become director.  On my first visit, he provided me with data to try
out a slight modification of the order selection procedure for a
vector autoregressive model-based ship autopilot, developed by him
in collaboration with marine engineers.  I~gave an invited
presentation on this work at the 1988 American Control Conference \cite{Fin88}.
My future GWU Ph.D. student Jim Cantor was in the audience, and was
stimulated to later ask me to direct his dissertation work on the
recursive estimation of incorrect time series models~\cite{CanFin06}.

I~also worked with Yoshinori Kawasaki at ISM, primarily in conjunction
with a series of conferences related to seasonal adjustment that
alternated venues between Tokyo and Washington, DC over a six-year
period.

At Academia Sinica, I~worked mostly with my host and co-author
Ching-Zong Wei, whose death from a rare disease ended a remarkable
research career much too soon.  I~also interacted with the
econometrician Jin-Lung Lin on various topics related to seasonal
adjustment.  He was at the Institute of Economics there, and is now
Dean of the College of Management of Taiwan's National Dong Hwa University in Hualien.

\begin{figure}

\includegraphics{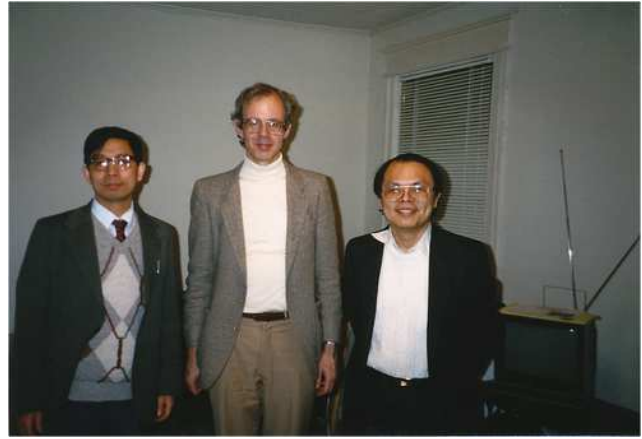}

\caption{Zhao-Guo Chen, David Findley and Ching-Zong
Wei at the home of David Findley, April 1989.}\label{f3}
\end{figure}

\textbf{McElroy}: Do you feel that the major goals of time series
research at USCB have been met?

\textbf{Findley}: The major goal has been to improve seasonal
adjustment practice at USCB and elsewhere.  This has involved the
development of new procedures, models, model selection procedures
and diagnostics.  Of course, one must evaluate these procedures for
efficacy and practicality.  Finally, it is necessary to implement
successful innovations in software.  Much of this work has proceeded
incrementally, in response to urgently perceived needs or
inadequacies with existing procedures or tools, or due to new
opportunities for collaboration with other seasonal adjusters.

It was always assumed that we would, in the course of time, develop
a new seasonal adjustment program and that we would make\vadjust{\goodbreak} greater use
of time series models, probably including the purely ARIMA
model-based seasonal adjustment procedure of Hillmer and Tiao \cite{HilTia82}.
George Tiao and his students have had a huge impact on the use of
models in seasonal adjustment.  When I~arrived at the Census Bureau in
1980, his students Steve Hillmer and Bill Bell were here.  Hillmer
was finishing his year as an ASA/NSF-Census Research Fellow with Bill Bell as his
Research Associate.  Bill was also finishing the writing of his
dissertation, which had path-breaking results on nonstationary
signal extraction relevant to model-based seasonal adjustment.  I~was able to hire Bill, and with occasional input and contributions
from me, he led the modeling developments for the next two decades
and more.  These efforts resulted in modeling innovations that were
incorporated into the X-12-ARIMA seasonal adjustment program
released by USCB in the 1990s \cite{BelHil84}.  This software is widely used around
the world.

Working with Brian Monsell and others, I~led the development of the
software's new seasonal adjustment quality diagnostics, as well as
model selection options concerned with transformations (such as
logarithms) and with nonnested choices among candidate regressors for
trading day and moving holiday effects.

The USCB plans to soon release the successor to X-12-ARIMA, which is
named X-13 ARIMA-SEATS.  Its new capabilities include the option to
produce Hillmer--Tiao model-based adjustments, and to compare them
with X-12-ARIMA adjustments using a common set of diagnostics.  This
should lead to a greater understanding of both traditional and
model-based adjustments.  The model-based adjustments will come from\vadjust{\goodbreak}
its incorporated version of the SEATS seasonal adjustment program,
thanks to the collaboration of its designer, Agustin Maravall of the
Bank of Spain.  Equally vital to this undertaking has been the
outstanding software development ability and statistical
understanding of Brian Monsell, who now leads the time series group.
He and Bill Bell were my most important hires.

\section{Past, Present and Future Research}

 \textbf{McElroy}:  Over your career you have made many contributions to
model selection and seasonal adjustment.  Of the different projects
and papers you have worked on, are there any favorites that you have or
that particularly stand out in your mind?

\textbf{Findley}: This is a question I~might give a different
answer to every time it is asked!  Three papers come to mind in model
selection.  I~have the feeling that the 1991 AISM paper \cite{Fin91N1} I~did on
counter-examples to BIC and other parsimonious model selection criteria
has some value.  It has been cited fairly often. What I~showed there
was that when you are working with models that are not correct, and you
are making nonnested model comparisons, that one unneeded parameter's
estimate can cause problems for forecasting or parameter estimation, in
general, that are significantly larger than the influence of another
parameter that is not needed.  In other words, we tend to think: if you
have $p$ unneeded parameters, then the distribution to describe this is
a $\chi^2_p$ distribution or something like this, and there is a kind
of homogeneity to the effects of these unneeded parameters; but it is
not true in the case of nonnested incorrect model comparisons.

Much of my research from early on has tried to address the situation of
incorrect models, and I~recommended the use of AIC at the Census
Bureau, because there is some justification and demonstrable practical
advantages for making comparisons between nonnested models with it.
This is something that comes up often in seasonal adjustment, and I~think in statistics in general.  Additionally, there aren't really
systematic ways of addressing nonnested model comparisons.

I'm also very pleased with the 2002 JMVA paper~\cite{FinWei02} I~wrote with Ching-Zong
Wei on a rigorous development of AIC for vector autoregressive models.
There were some powerful new mathematical results that others have
used, which were in that paper.  But also the paper contains a precise
theory of over-fitting,\vadjust{\goodbreak} which says that if you have these unneeded
parameters, they are going to corrupt the good results you might have
obtained from these models, using an independent replicate of the data.
In this situation, I~was able to develop a theory of over-fitting, so
that you can take the difference of the likelihood functions and use
that as some kind of measure of over-fit.  Mathematicians and
statisticians talk about over-fitting, but here was a case where I~could prove some results and come up with a rather precise measure---it
is pretty theoretical, but at least it's a handle on the concept of
over-fitting, making it a precise concept.

A third paper that comes to mind is the one I~just completed with
Tucker, published in the 2010 JSPI special issue honoring Manny Parzen \cite{McEFin10};
because, again, in one case it is another kind of likelihood ratio test
for nonnested model comparisons. That paper is generally concerned with
the problem of testing the forecasting performance of competing
models---that is, if you have a statistical test that involves saying,
``these two models are equally good at forecasting, one is not better
than the other,'' then you would like to have a test of some kind of
this hypothesis. We analyze an in-sample measure of forecast error, and
use as our null hypothesis that these two models forecast equally well,
and were able to derive a test statistic, in particular an asymptotic
distribution for the difference of the two forecast error
measures---and a consistent estimate of the variance of this asymptotic
distribution---which we could use to form a statistic you could use to
test whether one model is better than the other.

 I~had started on this topic, in the context of likelihood ratio tests, in the early
1980s.  I~had obtained an asymptotic distribution for the likelihood
ratio in the nonnested model case, having been inspired by the work of
the econometrician Quong Vuong, who had done similar things for
regression models, but I~had never been able to find a consistent
variance estimate for the asymptotic distribution, and Tucker 
and I~were able to obtain that in our paper.  Now, when we find time, we're
at work generalizing this to the vector autoregressive
situation.

Those are the papers I~like that are concerned with model selection.
For seasonal adjustment papers, I~particularly like papers that have
had good expository value.  I~was invited to write a paper that would
be the Journal of Business and Economics Statistics invited paper for
the 1996 Joint Statistical Meetings \cite{Finetal98}, and I~wrote that with Bill Bell
and other staff members---Brian Monsell, Bor-Chung Chen and Mark
Otto---and we really got a lot of ideas about seasonal adjustment and
the role of models and model selection, and even estimation.  It was a
discussion paper, and that brought up a lot of other ideas. So, I~regard that as a valuable paper.

I~would like to mention the 2006 Journal of Official Statistics paper I~did with Donald Martin on properties of time-domain filters, seasonal
adjustment filters applied to nonstationary series that really account
for the fact that we are working with a finite sample, and which
examined the effects of phase delay \cite{FinMar06}. I~am also very pleased with a
paper that Catherine Hood and I~wrote \cite{FinHOO00}, just setting out our procedures
for seasonally adjusting a bunch of time series we have never seen
before.  Specifically, how we go about choosing program settings for
them, or perhaps even deciding whether they should  be seasonally
adjusted at all.  The paper was written for a conference sponsored by
the Italian Statistical Office, where they gave out a dozen or so time
series and asked people to analyze them.  Subsequently, I~have been
very glad to have that paper available to hand out to people when I~have taught seasonal adjustment courses.

I~like the new trading day regressor that Brian Monsell and I~came up with, which is described in our 2009 Journal of Official
Statistics paper \cite{FinMon09}. It seems to do very well with stock series (e.g.,
inventories), better than the regressors we've had before.  Akaike
stressed to me the importance of models, and Bill Bell's concentration
on models from his education under George Tiao and George Box, has
certainly reinforced the power of models to me.  To convey statistical
information, and good statistical practice, models are an extremely
concentrated way of presenting information to people doing statistical
work.

\textbf{Holan}: In 1987 you were elected Fellow of the American
Statistical Association. Can you tell us a little bit about your award?

\textbf{Findley}: Yes, the award had particular meaning for me because,
while I~had had some coursework in stochastic processes in my Applied
Mathematics Ph.D. studies, I~had no formal training in statistics or
time series analysis. The award confirmed for me that I~had
successfully established myself as a statistician and time series
analyst. I~was delighted when Alan Izenman came up to me the day after
the award with a warm smile and handshake saying, ``Now no one can say
you aren't a statistician!''

\textbf{McElroy}: Well, you are now retired from your position at the
Census Bureau.  Are there future plans\vadjust{\goodbreak} for research, and are you still
active in research? What sorts of things are you doing?

\textbf{Findley}:  As I~mentioned, I~am interested in the
research you and I~are doing in extending our forecasting and
likelihood ratio tests to the vector situation, at least for vector
autoregressive models. I~should have mentioned that a special aspect of
this research is that the effect---at least when multi-step ahead
forecasting is involved---of parameter estimation is taken into account
in these tests, in a way that has not been done before with time series
models.  So I~think the paper's important for that reason, too---that
it is kind of a breakthrough.

I~hope someday we will find a way to understand what gain and phase
function graphs mean for nonstationary time series: I~think there is
some hope for this when only one differencing is involved in
transforming the data to stationarity.  But we haven't seen any reason
to be optimistic under more complicated differencing operations! I~have
some other papers on model selection that I~have never written up.  I~hope I'll get around to writing those up, in time---mostly having to do
with AIC, but for ARIMA models rather than AR models.  Also I~would
like to look at other situations, say where you're trying to decide
``should I~use a Weibull distribution or a logistic distribution to
model a certain kind of data?''  I~had a project on that, where I~got
some pretty formulas, but never quite the theory needed to justify the
formulas.

\textbf{Holan}: So, you've probably experienced a lot of changes in
statistics over the years.  What is your assessment of the state of
statistics in general and the future of seasonal adjustment in
particular?

\textbf{Findley}: I~don't feel qualified to comment on the state of
statistics in general.  But in terms of impact on statistics, the
internet has been the most important development!---for disseminating
data, for disseminating meta-data associated with the data, for
disseminating software, for disseminating the results of research, and
for maintaining contact with users.  Brian Monsell gets many messages
every month with seasonal adjustment questions; Tucker and I~get
messages from people we have taught seasonal adjustment courses to and
other places, that sometimes turn into research problems and papers.

In seasonal adjustment, with the release of X-13 ARIMA-SEATS, we
have just now gotten to the point where it is possible to compare
the results of a model-based seasonal adjustment and a seasonal
adjustment from the older X-11 filter methodologies. I~think that\vadjust{\goodbreak}
will be valuable and that, from having one software package, we will
learn more about both methods. We will be able to produce the
adjustments easily and at the same time produce some diagnostics
that make it easy to compare some aspects of the seasonal
adjustments.  Basically, you could say that if you run both methods
and they give pretty similar results, you should feel very good and
not think there is much to worry about; and if they produce rather
different results, it would be very good if you could come up with
some kind of understanding of why that is so.  I~think both methods
have different strengths and different weaknesses, and so I~think
that this software will help improve the practice of seasonal
adjustment.

I~also believe that there is room for further variations on the use of
these methods. When Akaike developed his BAYSEA program \cite{Aka80}, it essentially
forecasted a moving window of data; you could set it to take four or
five or more years of data, and it would produce a seasonal adjustment
of that set of data, and then move to another interval. Additionally,
it would take the center adjustment, from however many overlapping
spans, as being the final adjustment; so there would be two years of
data in this five year case, in which you would revise the seasonal
adjustment.  The econometrician David Hendry at Oxford, where I~visited
recently, is very keen on the idea that econometricians should, in many
instances, be using moving windows to do the modeling and analyses of
data that they want to do. Akaike did this in BAYSEA, that was a kind
of structural modeling setup, but I~think there is reason to utilize
this approach in other cases. You need more than five years of data to
estimate ARIMA coefficients, and perhaps seven or eight years of data
to estimate ARIMA coefficients well, and you also need about that much
data to estimate coefficients for trading day effects and moving
holidays like Easter, so there is some limitation on how small our
estimation window can be.  But maybe there are some tricks we don't
know about in terms of improving parameter estimates in smaller
samples.

I~think there are still some interesting options to be explored in
seasonal adjustment.  I~also think that the software that the
National Bank of Belgium is developing for the European Statistical
Office, which is intended to become an official seasonal adjustment
software package for the European Statistical Office, could be of
great benefit if it comes out to be as simple to modify and add to
as the plan of Jean Palate, the person directing its development at
the National Bank of Belgium.\vadjust{\goodbreak} New diagnostics and new variations on
ARIMA models or other models for seasonal adjustment could be
implemented quickly, and tested, and retained if practice shows them
to be valuable.

\textbf{McElroy}: On a different note, you have been in Washington, DC
many years now.  What was the intellectual culture like when you first
came here, and how has it changed?

\textbf{Findley}: I~think DC has for a very long time been an
intellectually exciting place in the general sense, because there is a
tremendous amount of scientific research done at the universities and
the large industrial consulting firms that work on military projects.
There is a large number of chapters of the Institute of Electronic
Engineers; the Washington Statistical Society is very large, and has
almost weekly seminars.  Specifically for seasonal adjustment, there
were more people active in seasonal adjustment research at the time
I~came here---at least outside the Census Bureau, that is to say. David
Pierce and Bill Cleveland were at the Federal Reserve Board; Bob
McIntyre was there along with some other people, as well as Stuart
Scott, Tom Evans and Dick Tiller at the Bureau of Labor Statistics
(BLS). So maybe things at BLS are like they were before.

 But to speak for the Census Bureau, I~think the
\mbox{recent} arrival of Bob Groves to be the director and the coming of
Rod Little to fill the newly restarted position of Associate
Director for Research and\break Methodology, will be very exciting. These
are\break world-class statisticians in their various branches of work,
both of whom have spent substantial amounts of time at the Census
Bureau before, and are going to be very supportive of research of
many different kinds, and very knowledgeable in terms of how they
make budget decisions regarding research.  I~believe our time series
staff is still quite strong, so I~am quite optimistic about the
continued stimulating intellectual atmosphere in the Census
Bureau.\looseness=1

\textbf{Holan}:  You brought up something interesting,\break name\-ly that
you developed a lot of software. Your background is mostly in
theoretical mathematics, and then you developed software along the way,
and did a lot of applied work.  I~think, in general, the field of
statistics has changed over the last thirty years, with the greater
emphasis on computing.  Also, certainly upon entering the workplace
today, one needs a different skill set than was the case twenty years
ago.  What do you see are the important things for people to learn in
school, in order to be successful in this environment?\vadjust{\goodbreak}

\textbf{Findley}: I~think I~can only speak to what the Census
Bureau would like to see in young statisticians: some training in time
series analysis, and some knowledge of a programming language---it
could be C or something else, but more than just knowledge of a
statistical package like SAS.  Knowledge of R would be a great asset,
because that is probably the most important proto-typing language in
statistics now---I~certainly think that is true for time series
analysis. It would be great if people knew some sampling methodology
too, which is taught irregularly at universities, as is time series.
Experimental design would be a good thing.  We're always looking for
people who can work on experimental design projects that come up at the
Census Bureau in various contexts.

\textbf{McElroy}:  Well, it's been good talking to you. I~know that
some time ago you had a Statistical Science Conversation with Hirotugu Akaike, that you
conducted in collaboration with Manny Parzen \cite{FinPar95}.  Can you tell us a
little about that interview?

\textbf{Findley}:  It was wonderful to be able to work with
Manny.  It was my idea to have the interview, but Manny agreed to
work with me as soon as I~asked him.  Manny was the person in the
United States who discovered Akaike, who became aware of the work
Akaike was publishing in Japanese journals, and of its very high
quality and considerable interest. This is discussed in the
interview, so I~think the interview was substantially richer for
Manny's participation in it.  It was a delight to have this kind of
collaboration with Manny, whom I've known for a long time, and asked
for assistance on certain other projects, but never done any joint
research with.

 I~learned some interesting things from the
interview.  The statement about models being an extremely compressed
and portable form of information, is something Akaike said in the
interview.  Another thing he said, was that he didn't like vector
autoregressive models, at least initially, because it's so hard to
make sense of all the coefficients.  But, he discovered they were
extremely powerful in applications, so he decided to like them!  The
other thing he said was that a visit he made to Princeton, where
John Tukey was, at Tukey's invitation, was very influential for him.
He had been very theoretically based, his doctorate was in
mathematics from the University of Tokyo, and he had done some
reading in statistics.  However, Akaike said at Princeton he saw
Tukey give a lecture on robust methods and outliers.  Tukey, who was
revered as a mathematician (e.g.,  Tukey's Lemma in topology among
other contributions) presented methods that seemed very sensible, but
which lacked a mathematical foundation.  It was very helpful for
Akaike to see that this was acceptable; that is, that it seemed a
sensible way to practice statistics.  I~have felt that, if you can't
find an appropriate theory to justify what you're doing, but it
seems to be producing sensible results, don't give up on the search
for a stronger foundation, but don't give up on the method
either!\looseness=1

\textbf{McElroy}: It seems to be a good motto! Anything else you would
like to add?

\textbf{Findley}: It's been very important to whatever success
I've had at USCB and elsewhere, to have really good colleagues to work
with.  I~think most of us benefit from interactions with colleagues
with different backgrounds from ours, but particularly when you are
doing applied work and when you are trying to produce software that
other people can use, you have to have the right people to work with.
It also helps when you are trying to solve a very difficult theoretical
problem, like some of the problems I~have worked on with Ching-Zong
Wei!  It's certainly important to have some luck.  Bill Bell was here
when I~came, Brian Monsell showed up afterward, and there have been
other excellent people that came and left over the years.  I~was lucky
that the mathematics department at the University of Tulsa had its
particular composition of mathematicians who knew you could become a
time series analyst without formal training. But I~have to say, I~am
very grateful for the opportunities the Census Bureau has given me.  I~have always gotten good support for my work, even things like
permission to go off for months to some other place!  So it has been a
very good place to work, and I~have no reason to think that the
Statistical Research Division isn't just as good a place now to work as
it was before, when I~was here full time.

\textbf{Holan}: Well, thank you for giving us this opportunity to talk
with you today.

\textbf{Findley}: Sure.

\textbf{McElroy}: Thank you very much.

\section*{Acknowledgments}
The authors would like to thank the Editor, Jon Wellner, for providing
a suggestion that helped improve this article. Holan's research was
partially supported through the ``Summer at Census'' research program
(2010).

\section*{Disclaimer}
This paper is released to inform interested parties of ongoing research
and to encourage discussion of work in progress.  The views expressed
are those of the authors and not necessarily those of the U.S. Census
Bureau.



\begin{thebibliography}{27}

\bibitem{Aka80}
\begin{barticle}[mr]
\bauthor{\bsnm{Akaike},~\bfnm{Hirotugu}\binits{H.}}
(\byear{1980}).
\btitle{Seasonal adjustment by a {B}ayesian modeling}.
\bjournal{J. Time Ser. Anal.}
\bvolume{1}
\bpages{1--13}.
\bid{doi={10.1111/j.1467-9892.1980.tb00296.x}, issn={0143-9782}, mr={0605571}}
\bptok{imsref}%
\end{barticle}
\endbibitem

\bibitem{And71}
\begin{bbook}[mr]
\bauthor{\bsnm{Anderson},~\bfnm{T.~W.}\binits{T.~W.}}
(\byear{1971}).
\btitle{The Statistical Analysis of Time Series}.
\bpublisher{Wiley}, \baddress{New York}.
\bid{mr={0283939}}
\bptok{imsref}%
\end{bbook}
\endbibitem

\bibitem{BelHil84}
\begin{barticle}[auto:STB|2012/06/08|12:49:54]
\bauthor{\bsnm{Bell},~\bfnm{W.}\binits{W.}} \AND
  \bauthor{\bsnm{Hillmer},~\bfnm{S.}\binits{S.}}
(\byear{1984}).
\btitle{Issues involved with the seasonal adjustment of economic time series}.
\bjournal{J.~Bus. Econom. Statist.}
\bvolume{2}
\bpages{291--320}.
\bptok{imsref}%
\end{barticle}
\endbibitem

\bibitem{BoxJen76}
\begin{bbook}[auto:STB|2012/06/08|12:49:54]
\bauthor{\bsnm{Box},~\bfnm{G.}\binits{G.}} \AND
  \bauthor{\bsnm{Jenkins},~\bfnm{G.}\binits{G.}}
(\byear{1976}).
\btitle{Time Series Analysis}.
\bpublisher{Holden-Day}, \baddress{San Francisco}.
\bptok{imsref}%
\end{bbook}
\endbibitem

\bibitem{Bri01}
\begin{bbook}[mr]
\bauthor{\bsnm{Brillinger},~\bfnm{David~R.}\binits{D.~R.}}
(\byear{1975}).
\btitle{Time Series: Data Analysis and Theory}.
\bpublisher{SIAM}, \baddress{Philadelphia, PA}.
\bptnote{check year}%
\bptok{imsref}%
\end{bbook}
\endbibitem

\bibitem{CanFin06}
\begin{bincollection}[mr]
\bauthor{\bsnm{Cantor},~\bfnm{James~L.}\binits{J.~L.}} \AND
  \bauthor{\bsnm{Findley},~\bfnm{David~F.}\binits{D.~F.}}
(\byear{2006}).
\btitle{Recursive estimation of possibly misspecified MA(1) models: Convergence
  of a general algorithm}.
In \bbooktitle{Time Series and Related Topics},
(\beditor{\bfnm{H.~C.}\binits{H.~C.}~\bsnm{Ho}},
  \beditor{\bfnm{C.~K.}\binits{C.~K.}~\bsnm{Ing}} \AND
  \beditor{\bfnm{T.~L.}\binits{T.~L.}~\bsnm{Lai}}, eds.).
\bseries{Institute of Mathematical Statistics Lecture Notes---Monograph Series}
\bvolume{52}
\bpages{20--47}.
\bpublisher{IMS}, \baddress{Beachwood, OH}.
\bid{doi={10.1214/074921706000000932}, mr={2427837}}
\bptnote{check year}%
\bptok{imsref}%
\end{bincollection}
\endbibitem

\bibitem{Dur60}
\begin{barticle}[auto:STB|2012/06/08|12:49:54]
\bauthor{\bsnm{Durbin},~\bfnm{J.}\binits{J.}}
(\byear{1960}).
\btitle{The fitting of time-series models}.
\bjournal{Revue, Institut International de Statistique}
\bvolume{28}
\bpages{233--243}.
\bptok{imsref}%
\end{barticle}
\endbibitem

\bibitem{autokey10}
\begin{bbook}[mr]
\beditor{\bsnm{Findley},~\bfnm{David}\binits{D.}}, ed.
(\byear{1978}).
\btitle{Applied Time Series Analysis}.
\bpublisher{Academic Press}, \baddress{New York}.
\bid{mr={0507851}}
\bptok{imsref}%
\end{bbook}
\endbibitem

\bibitem{Fin81}
\begin{bincollection}[mr]
\bauthor{\bsnm{Findley},~\bfnm{David}\binits{D.}}
(\byear{1981}).
\btitle{Geometrical and lattice versions of {L}evinson's general algorithm}.
In \bbooktitle{Applied Time Series Analysis, {II} ({T}ulsa, {O}kla., 1980)}
\bpages{327--354}.
\bpublisher{Academic Press}, \baddress{New York}.
\bid{mr={0651942}}
\bptok{imsref}%
\end{bincollection}
\endbibitem

\bibitem{autokey11}
\begin{bbook}[mr]
\beditor{\bsnm{Findley},~\bfnm{David}\binits{D.}}, ed.
(\byear{1981}).
\btitle{Applied Time Series Analysis. {II}}.
\bpublisher{Academic Press}, \baddress{New York}.
\bid{mr={0651933}}
\bptok{imsref}%
\end{bbook}
\endbibitem

\bibitem{Fin88}
\begin{bincollection}[auto:STB|2012/06/08|12:49:54]
\bauthor{\bsnm{Findley},~\bfnm{D.}\binits{D.}}
(\byear{1988}).
\btitle{An analysis of AIC for linear stochastic regression and control}.
In \bbooktitle{Proceedings of the 1988 American Control Conference}
\bpages{1281--1288}.
\bptok{imsref}%
\end{bincollection}
\endbibitem

\bibitem{Fin91N1}
\begin{barticle}[mr]
\bauthor{\bsnm{Findley},~\bfnm{David}\binits{D.}}
(\byear{1991}).
\btitle{Counterexamples to parsimony and {BIC}}.
\bjournal{Ann. Inst. Statist. Math.}
\bvolume{43}
\bpages{505--514}.
\bid{doi={10.1007/BF00053369}, issn={0020-3157}, mr={1143638}}
\bptok{imsref}%
\end{barticle}
\endbibitem

\bibitem{Fin91N2}
\begin{barticle}[mr]
\bauthor{\bsnm{Findley},~\bfnm{David}\binits{D.}}
(\byear{1991}).
\btitle{Convergence of finite multistep predictors from incorrect models and
  its role in model selection}.
\bjournal{Note Mat.}
\bvolume{11}
\bpages{145--155}.
\bid{issn={1123-2536}, mr={1258544}}
\bptnote{check year}%
\bptok{imsref}%
\end{barticle}
\endbibitem

\bibitem{FinHOO00}
\begin{barticle}[auto:STB|2012/06/08|12:49:54]
\bauthor{\bsnm{Findley},~\bfnm{D.}\binits{D.}} \AND
  \bauthor{\bsnm{HOOD},~\bfnm{C.}\binits{C.}}
(\byear{2000}).
\btitle{X-12-ARIMA and its application to some Italian indicator series. In
  ``Seasonal Adjustment Procedures -- Experiences and Perspectives.''}.
\bjournal{Annali di Statistica Anno 129 Serie X}
\bvolume{20}
\bpages{249--269}.
\bptok{imsref}%
\end{barticle}
\endbibitem

\bibitem{FinMar06}
\begin{barticle}[auto:STB|2012/06/08|12:49:54]
\bauthor{\bsnm{Findley},~\bfnm{D.}\binits{D.}} \AND
  \bauthor{\bsnm{Martin},~\bfnm{D.}\binits{D.}}
(\byear{2006}).
\btitle{Frequency domain analyses of SEATS and X-11/12-ARIMA seasonal
  adjustment filters for short and moderate-length time series}.
\bjournal{Journal of Official Statistics}
\bvolume{22}
\bpages{1--34}.
\bptok{imsref}%
\end{barticle}
\endbibitem

\bibitem{FinMon09}
\begin{barticle}[auto:STB|2012/06/08|12:49:54]
\bauthor{\bsnm{Findley},~\bfnm{D.}\binits{D.}} \AND
  \bauthor{\bsnm{Monsell},~\bfnm{B.}\binits{B.}}
(\byear{2009}).
\btitle{Modeling stock trading day effects under flow day-of-week constraints}.
\bjournal{Journal of Official Statistics}
\bvolume{25}
\bpages{415--430}.
\bptok{imsref}%
\end{barticle}
\endbibitem

\bibitem{Finetal98}
\begin{barticle}[auto:STB|2012/06/08|12:49:54]
\bauthor{\bsnm{Findley},~\bfnm{D.}\binits{D.}},
  \bauthor{\bsnm{Monsell},~\bfnm{B.}\binits{B.}},
  \bauthor{\bsnm{Bell},~\bfnm{W.}\binits{W.}},
  \bauthor{\bsnm{Otto},~\bfnm{M.}\binits{M.}} \AND
  \bauthor{\bsnm{Chen},~\bfnm{B.}\binits{B.}}
(\byear{1998}).
\btitle{New capabilities and methods of the X-12-ARIMA seasonal adjustment
  program}.
\bjournal{J.~Bus. Econom. Statist.}
\bvolume{16}
\bpages{127--177}.
\bptok{imsref}%
\end{barticle}
\endbibitem

\bibitem{FinPar95}
\begin{barticle}[auto:STB|2012/06/08|12:49:54]
\bauthor{\bsnm{Findley},~\bfnm{D.}\binits{D.}} \AND
  \bauthor{\bsnm{Parzen},~\bfnm{E.}\binits{E.}}
(\byear{1995}).
\btitle{A conversation with Hirotugu Akaike}.
\bjournal{Statist. Sci.}
\bvolume{10}
\bpages{104--117}.
\bptok{imsref}%
\end{barticle}
\endbibitem

\bibitem{FinPotWei01}
\begin{barticle}[mr]
\bauthor{\bsnm{Findley},~\bfnm{David}\binits{D.}},
  \bauthor{\bsnm{P{\"o}tscher},~\bfnm{Benedikt~M.}\binits{B.~M.}} \AND
  \bauthor{\bsnm{Wei},~\bfnm{Ching-Zong}\binits{C.-Z.}}
(\byear{2001}).
\btitle{Uniform convergence of sample second moments of families of time series
  arrays}.
\bjournal{Ann. Statist.}
\bvolume{29}
\bpages{815--838}.
\bid{doi={10.1214/aos/1009210691}, issn={0090-5364}, mr={1865342}}
\bptok{imsref}%
\end{barticle}
\endbibitem

\bibitem{FinPotWei04}
\begin{barticle}[mr]
\bauthor{\bsnm{Findley},~\bfnm{David}\binits{D.}},
  \bauthor{\bsnm{P{\"o}tscher},~\bfnm{Benedikt~M.}\binits{B.~M.}} \AND
  \bauthor{\bsnm{Wei},~\bfnm{Ching-Zong}\binits{C.-Z.}}
(\byear{2004}).
\btitle{Modeling of time series arrays by multistep prediction or likelihood
  methods}.
\bjournal{J. Econometrics}
\bvolume{118}
\bpages{151--187}.
\bid{doi={10.1016/S0304-4076(03)00139-8}, issn={0304-4076}, mr={2030971}}
\bptok{imsref}%
\end{barticle}
\endbibitem

\bibitem{FinWei02}
\begin{barticle}[mr]
\bauthor{\bsnm{Findley},~\bfnm{David}\binits{D.}} \AND
  \bauthor{\bsnm{Wei},~\bfnm{Ching-Zong}\binits{C.-Z.}}
(\byear{2002}).
\btitle{A{IC}, overfitting principles, and the boundedness of moments of
  inverse matrices for vector autoregressions and related models}.
\bjournal{J. Multivariate Anal.}
\bvolume{83}
\bpages{415--450}.
\bid{doi={10.1006/jmva.2001.2063}, issn={0047-259X}, mr={1945962}}
\bptok{imsref}%
\end{barticle}
\endbibitem

\bibitem{HilTia82}
\begin{barticle}[mr]
\bauthor{\bsnm{Hillmer},~\bfnm{C.}\binits{C.}} \AND
  \bauthor{\bsnm{Tiao},~\bfnm{G.~C.}\binits{G.~C.}}
(\byear{1982}).
\btitle{An {ARIMA}-model-based approach to seasonal adjustment}.
\bjournal{J.~Amer. Statist. Assoc.}
\bvolume{77}
\bpages{63--70}.
\bid{issn={0162-1459}, mr={0648026}}
\bptok{imsref}%
\end{barticle}
\endbibitem

\bibitem{Lev47}
\begin{barticle}[mr]
\bauthor{\bsnm{Levinson},~\bfnm{Norman}\binits{N.}}
(\byear{1947}).
\btitle{The {W}iener {RMS} (root mean square) error criterion in filter design
  and prediction}.
\bjournal{J. Math. Phys.}
\bvolume{25}
\bpages{261--278}.
\bid{issn={0097-1421}, mr={0019257}}
\bptnote{check year}%
\bptok{imsref}%
\end{barticle}
\endbibitem

\bibitem{McEFin10}
\begin{barticle}[mr]
\bauthor{\bsnm{McElroy},~\bfnm{Tucker~S.}\binits{T.~S.}} \AND
  \bauthor{\bsnm{Findley},~\bfnm{David~F.}\binits{D.~F.}}
(\byear{2010}).
\btitle{Selection between models through multi-step-ahead forecasting}.
\bjournal{J. Statist. Plann. Inference}
\bvolume{140}
\bpages{3655--3675}.
\bid{doi={10.1016/j.jspi.2010.04.032}, issn={0378-3758}, mr={2674155}}
\bptok{imsref}%
\end{barticle}
\endbibitem

\bibitem{RiNa55}
\begin{bbook}[auto]
\bauthor{\bsnm{Riesz},~\bfnm{F.}\binits{F.}} \AND
  \bauthor{\bsnm{Sz.-Nagy},~\bfnm{B.}\binits{B.}}
(\byear{1955}).
\btitle{Functional Analysis}.
\bpublisher{Unger}, \baddress{New York}.
\bptok{imsref}%
\end{bbook}
\endbibitem

\bibitem{Rob80}
\begin{bbook}[mr]
\bauthor{\bsnm{Robinson},~\bfnm{Enders~A.}\binits{E.~A.}}
(\byear{1980}).
\btitle{Physical Applications of Stationary Time-Series}.
\bpublisher{Macmillan Inc.}, \baddress{New York}.
\bid{mr={0609704}}
\bptok{imsref}%
\end{bbook}
\endbibitem

\end{thebibliography}
\end{document}